\documentclass[12pt, preprint]{aastex}
\usepackage{amsmath}
\usepackage{graphicx}
\shorttitle{Calibrating convection in stars}
\shortauthors{Bonaca et al.}

\begin{document}
\title{Calibrating Convective properties of Solar-like Stars in the {\it Kepler} Field of View}

\author{
Ana~Bonaca\altaffilmark{1},
Joel~D.~Tanner\altaffilmark{1},
Sarbani~Basu\altaffilmark{1,2},
William~J.~Chaplin\altaffilmark{3,2},
Travis~S.~Metcalfe\altaffilmark{4,2},
M\'ario~J.~P.~F.~G.~Monteiro\altaffilmark{5},
J\'er\^ome Ballot\altaffilmark{6},
Timothy~R.~Bedding\altaffilmark{7},
Alfio~Bonanno\altaffilmark{8},
Anne-Marie~Broomhall\altaffilmark{3},
Hans~Bruntt\altaffilmark{9},
Tiago L. Campante\altaffilmark{5},
J{\o}rgen~Christensen-Dalsgaard\altaffilmark{9,2},
Enrico~Corsaro\altaffilmark{7},
Yvonne~Elsworth\altaffilmark{3},
Rafael~A.~Garc\'{\i}a\altaffilmark{10,2},
Saskia~Hekker\altaffilmark{11},
Christoffer~Karoff\altaffilmark{9},
Hans~Kjeldsen\altaffilmark{9},
Savita~Mathur\altaffilmark{4,2},
Clara~R\'egulo\altaffilmark{12,13},
Ian~Roxburgh\altaffilmark{14},
Dennis~Stello\altaffilmark{7},
Regner Trampedach\altaffilmark{15},
Thomas~Barclay\altaffilmark{16},
Christopher~J.~Burke\altaffilmark{17},
Douglas~A.~Caldwell\altaffilmark{17}
}

\altaffiltext{1}{Department of Astronomy, Yale University, PO Box 208101, New Haven, CT 06520-8101; ana.bonaca@yale.edu,
sarbani.basu@yale.edu, joel.tanner@yale.edu}
\altaffiltext{2}{Kavli Institute for Theoretical Physics, Kohn Hall, University of California, Santa Barbara, CA 93106}
\altaffiltext{3}{School of Physics and Astronomy, University of
Birmingham, Edgbaston, Birmingham B15 2TT, UK}
\altaffiltext{4}{High Altitude Observatory, NCAR, P.O. Box 3000, Boulder CO 80301}
\altaffiltext{5}{Centro de Astrof\'{\i}sica and Faculdade de Ci\^encias, Universidade do Porto,
     Rua das Estrelas, 4150-762 Porto, Portugal}
\altaffiltext{6}{CNRS, Institut de Recherche en Astrophysique et Plan\'etologie, 14 avenue Edouard Belin, 31400 Toulouse, France;
Universit\'e de Toulouse, UPS-OMP, IRAP, Toulouse, France}
\altaffiltext{7}{Sydney Institute for Astronomy (SIfA), School of Physics, University of Sydney, NSW 2006, Australia}
\altaffiltext{8}{INAF - Osservatorio Astrofisico di Catania, Via S.Sofia 78, 95123 Catania, Italy}
\altaffiltext{9}{Stellar Astrophysics Centre, Department of Physics and Astronomy, Aarhus University, DK-8000 Aarhus C, Denmark}
\altaffiltext{10}{Laboratoire AIM, CEA/DSM-CNRS-Universiti\'e Paris Diderot; IRFU/SAp, Centre de Saclay, 91191 Gif-sur-Yvette
Cedex, France}
\altaffiltext{11}{Astronomical Institute “Anton Pannekoek”, University of Amsterdam, Science Park 904, 1098 XH Amsterdam, The Netherlands}
\altaffiltext{12}{Instituto de Astrofísica de Canarias, 38205, La Laguna, Tenerife, Spain}
\altaffiltext{13}{Universidad de La Laguna, Dpto de Astrofísica, 38206, La Laguna, Tenerife, Spain}
\altaffiltext{14}{School of Physics and Astronomy, Queen Mary University of London, Mile End Road, London E1 4NS, UK}
\altaffiltext{15}{JILA, University of Colorado and National Institute of Standards and
    Technology, 440 UCB, Boulder, CO 80309}
\altaffiltext{16}{Bay Area Environmental Research Inst./NASA Ames Research Center, Moffett Field, CA 94035}
\altaffiltext{17}{SETI Institute/NASA Ames Research Center, Moffett Field, CA 94035}

\begin{abstract}
Stellar models generally use simple parametrizations to treat convection. The most
widely used parametrization is the so-called ``Mixing Length Theory'' where the convective eddy sizes are
described using a single number, $\alpha$, the  mixing-length parameter. This
is a free parameter, and the general practice is to calibrate $\alpha$ using the known properties
of the Sun and apply that to all stars.
Using data from NASA's {\it Kepler} mission we
show that using the solar-calibrated  $\alpha$ is not always appropriate, and that
in many cases it would lead to estimates of initial helium abundances that are lower
than the primordial helium abundance. {\it Kepler} data allow us to calibrate $\alpha$
for many other stars and we show that for the sample of stars we have studied,
the mixing-length parameter is generally lower than the solar value. 
We studied the correlation between $\alpha$ and stellar properties, and we find
that $\alpha$ increases with metallicity. 
We therefore conclude that results obtained by fitting stellar models or by using
population-synthesis models  constructed with solar values of $\alpha$ are likely to have
large systematic errors.
Our results also confirm theoretical expectations that the mixing-length parameter
should vary with stellar properties.
\end{abstract}
\maketitle

\section{Introduction}
\label{intro}

Accurately treating convective heat transport in stellar models is difficult.
The structure and evolution of most stars is related to
convective transport processes  in their outer layers.
The transition from efficient convective transport in the deep envelope to the
radiative atmospheric layers takes place in a region of inefficient
convection where the temperature gradient is highly superadiabatic. The
poorly known structure of this region remains one of the major uncertainties in
stellar models. In one-dimensional calculations, this region 
is usually modeled using the ``mixing length theory,'' or MLT
\citep{boh58}. This prescription assumes that one can approximate
the full range of turbulent  eddy-sizes by a typical
size, and that an eddy on average travels a distance  determined by the eddy size
before losing its identity. This distance, is
known as the ``mixing length'' and is usually  defined as $\alpha H_p$, where $\alpha$ is
known as the ``mixing-length parameter,'' 
and $H_p$ is the local pressure scale height.  Given the atmospheric structure, the approximation, when
given the mixing length, fix the specific entropy in
the deep convection zone, which in turn determines the radius of the model.
{ The model radius thus depends sensitively on the choice of the mixing-length parameter
$\alpha$, which is a free parameter that cannot be determined from the 
mixing length theory.}

The common practice when modeling stars to determine their structure and evolution
is to use the solar value of $\alpha$. We know the mass, radius,
luminosity and age of the Sun precisely. Solar models are constructed by searching for $\alpha$ and the initial
helium abundance $Y_0$ that yield a model with the correct radius and luminosity 
at the Sun's age. Since masses, radii and  ages are
usually unknown for other stars, the solar approach is unfeasible and, hence, the solar
value of $\alpha$ is used.

The assumption of a fixed $\alpha$ has no {\it a priori} justification, and indeed,
there is some evidence that the solar value of $\alpha$ does not always work for other
stars. \citet{la84} first noted that
the radius of $\alpha$~Cen~A cannot be reproduced using the solar value of 
$\alpha$. A combined astrometric and seismic study of the
$\alpha$~Cen system by \citet{de86} confirmed this result as have more
recent studies (e.g., Fernandes \& Neuforge 1995; Miglio \& Montalban 2005).
Some studies of other
binary systems have also suggested a mass-dependence of $\alpha$
(e.g., Ludwig \& Salaris 1999; Morel et al. 2000; Lebreton et al. 2001;
Lastennet et al. 2003, etc.). In particular,  Y{\i}ld{\i}z et al.~(2006) studied
binaries in the Hyades cluster and suggested that $\alpha$ increases with stellar mass.
Numerical simulations of stellar convection also suggest that convective properties
vary with stellar parameters (e.g., Ludwig et al. 1999; Trampedach 2007; Trampedach \& Stein
2011).

While studies of binaries and clusters have suggested  that the solar $\alpha$
is not always applicable, the situation for single field stars is not clear
because of the lack of observational constraints. 
Asteroseismic data from space missions like CoRoT (Michel et al. 2008)
 and {\it Kepler} (Borucki et al. 2010) allow us to
place independent constraints on the mass and radius of single stars. These constraints,
along with the classical constraints of $T_{\rm eff}$ and metallicity, allow us to
constrain $\alpha$. We already know that the oscillation spectra of some CoRoT 
and {\it Kepler} stars cannot be reproduced using the solar value of $\alpha$ (e.g., Metcalfe
et al. 2010; Deheuvels \& Michel 2011, Mathur et al. 2012). 
Detailed asteroseismic modeling %
of the solar analogs 16 Cyg A \& B has also required the adoption of non-solar values of $\alpha$ %
(Metcalfe et al. 2012). %

In this study we use asteroseismic data obtained by NASA's {\it Kepler} mission to
calibrate $\alpha$ for a sample of dwarfs and subgiants.
Observational data and our analysis technique are described
in \S~\ref{sec:methods}.  We present our results and discuss their implications in \S~\ref{sec:results}.

It should be noted that $\alpha$ is just a proxy for describing stellar convection, and
in particular $\alpha$ describes  the entropy change between the surface and the deeper, isentropic, layers 
of efficient convection.  As a result,
the value of $\alpha$ cannot be derived uniquely. It depends on 
the exact formulation of the mixing-length theory (see, e.g., Appendix A of Ludwig et al. 1999), as well as
physics inputs  that affect entropy e.g., atmospheric opacities, the temperature-optical depth relation
($T$-$\tau$) in the atmosphere, the equation of state and processes such as 
the gravitational settling of helium
and heavy elements.  Thus, the value of $\alpha$ in a given star needs to be examined in the context of 
the value of $\alpha$ needed to construct a solar model with the same physics. 

{  Note that conventional
MLT assumes a fixed ratio between the distance traveled by an eddy and its size,
and $\alpha$ is the only free parameter.  Some approximations (see e.g., Arnett et al. 2010) leave the
ratio as an adjustable parameter. }
We use the conventional B\"ohm-Virtense form of MLT and only adjust $\alpha$.

\section{Method}
\label{sec:methods}

We used data obtained by the {\it Kepler} asteroseismology program (Gilliland et al. 2010) during
its survey phase. 
The survey collected data on more than 2000 main-sequence and sub-giant stars and stellar oscillations were detected in about
500 (Chaplin et al. 2011).  Of these we used a subset of 90 stars for which spectroscopic estimates
of effective temperature $T_{\rm eff}$ and metallicity [M/H] are available from
Bruntt et al. (2012).

We used the average large separation $\Delta\nu$ and the frequency of maximum oscillation power $\nu_{\rm max}$ for this work.
The large separation scales approximately as the square root of the mean density of a star (Ulrich 1986; Christensen-Dalsgaard 1988)
while the frequency of maximum power scales approximately as $g/\sqrt{T_{\rm eff}}$ (Brown et al. 1991; 
Kjeldsen \& Bedding 1995; Bedding \& Kjeldsen 2003).
The $\Delta\nu$ and $\nu_{\rm max}$ values  are those used by Verner et al. (2011) to verify the
{\it Kepler} Input Catalog (Brown et al. 2011).

First, $\Delta\nu$, $\nu_{\rm max}$, $T_{\rm eff}$ and [M/H] are used to estimate the mass and radius of
each star. For this we used the grid-based Yale-Birmingham (YB) pipeline described
by  Basu et al.~(2010) Gai et al.~(2011) and Basu et al.~(2012).
We use four stellar-model grids for our work: models from the Yonsei-Yale (YY) isochrones (Demarque et al. 2004), and those
of Dotter et al. (2008), Marigo et al. (2008) and Gai et al. (2010).
The grids have been constructed with different physics inputs and modeling parameters, and with solar $\alpha$ consistent
with the inputs to the grids.
For subsequent calculations the average of the mass and radius estimates returned by the four grids is used. 

\begin{figure*}
\epsscale{0.47}
\plotone{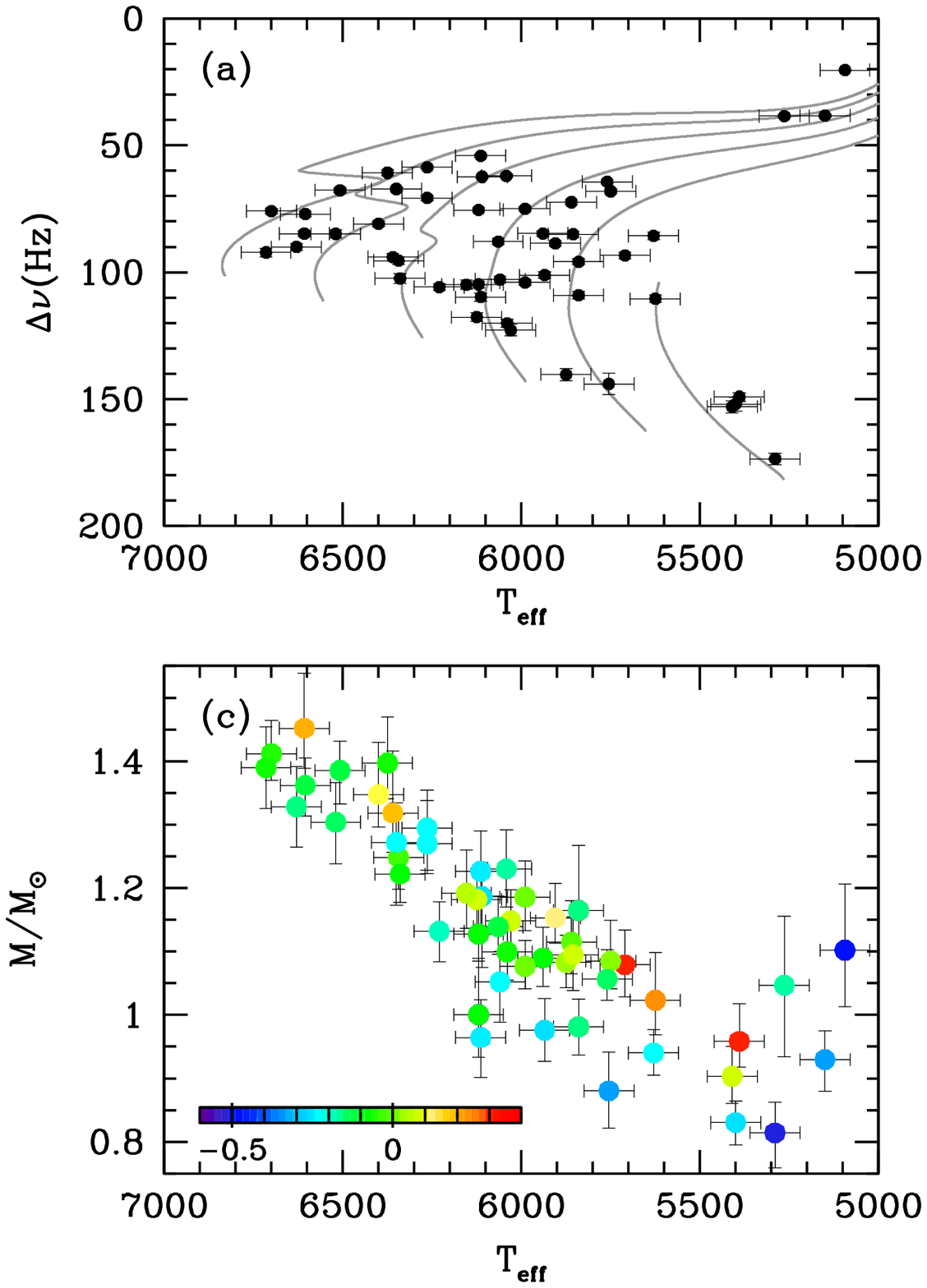}
\plotone{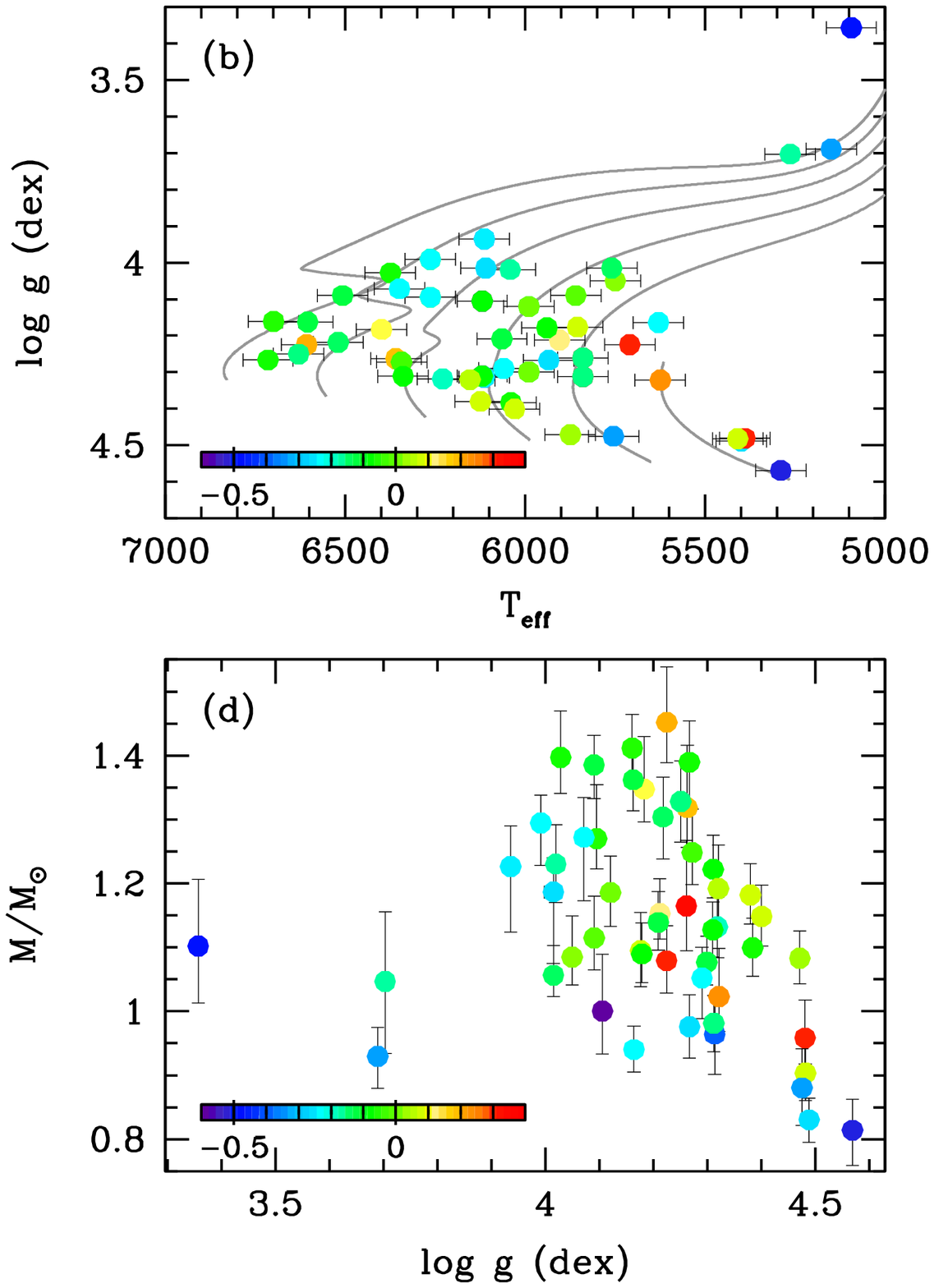}
\caption{(a) $\Delta\nu$ of the stars plotted against $T_{\rm eff}$. The  gray lines are tracks for
masses 0.9 M$_\odot$ to 1.4 M$_\odot$; (b) Derived $\log g$ plotted against $T_{\rm eff}$; (c) The derived
mass as a function of $T_{\rm eff}$; and (d) the derived mass plotted as a function of the derived $\log g$.
In panels (b)-(d) the color scale denotes metallicity (dex).}
\label{fig:props}
\end{figure*}

The properties of
our final sample are shown in Fig.~\ref{fig:props}. Note that our sample
is quite restricted in terms of $\log g$ and  we lack stars that are close to the base of the red giant branch; in
fact we have very few stars with $\log g < 4$. 
This will make finding a $\log g$ dependence of the derived $\alpha$ difficult. There is the expected correlation
between mass and $T_{\rm eff}$ --- the higher temperature stars are generally  more massive. Although 
each mass range spans a range of $\log g$, unsurprisingly the least evolved stars in our sample are also the least massive.
Most  stars in our sample have sub-solar metallicities.

Starting with mass, radius, $T_{\rm eff}$ and metallicity, each star in our
sample was modeled using the Yale Stellar Evolution Code (YREC; Demarque et al. 2008) in an iterative manner.
In this mode, radius and $T_{\rm eff}$ were specified and the code determined either $\alpha$ or
the initial helium abundance ($Y_0$) that yielded the specified radius at the given $T_{\rm eff}$
for a given mass. In the former case $Y_0$ has to be specified, in the latter $\alpha$
has to be specified. 

The input physics consisted of  the OPAL equation of state (Rogers \& Nayfonov 2002), 
OPAL high-temperature opacities (Iglesias \& Rogers 1996) 
supplemented with Ferguson et al.~(2005) low temperature opacities.
Nuclear reaction rates were from Adelberger et al.~(1998), except for the 
 $^{14}N(p,\gamma)^{15}O$ reaction, which was fixed at the value of 
Formicola et al.~(2004). { Models did not include core overshoot or the
diffusion and settling of helium and heavy elements. }
We used the Eddington $T-\tau$ relation in the atmosphere.
With the above physics, the  solar-calibrated value of  $\alpha$ is $1.690$. 
Including  gravitational settling of helium and metals would change that
to $1.826$.

For the first set of calculations we assumed solar $\alpha$ for all stars and determined $Y_0$ that would be needed to
model the stars. We then performed two other sets of calculations: (1) we estimated $\alpha$  assuming that all stars have 
the solar value of $Y_0=0.278$ (Serenelli \& Basu 2010); (2) we estimated $\alpha$  assuming that $Y_0$ follows a simple 
chemical evolution model, $Y_0= 0.245+1.54Z$ (Dotter et al. 2008).

The iterative modeling process was repeated for 200 Monte Carlo
realizations of $R,\; M,\; T_{\rm eff}\;$ and [M/H] to estimate the uncertainties in $Y_0$ (or $\alpha$). Parameters in each 
realization were randomly chosen from a Gaussian distribution centered 
on the measured central value, with the dispersion equal to the mean measurement error. To avoid uncertainties due to 
small number statistics, we show  results of stars with at least 20 converged iterations.
This  requirement resulted in a final sample of about 55 stars.
The median of the distribution of 
parameter values is quoted as the central value of the parameter, and uncertainties are determined as
the 68\% confidence limit of the distribution. 

The reliability of the derived  $\alpha$ estimates  depends on the reliability of the observations and
those of the  mass and radius estimates.  There are indications that grid-based mass  estimates  
can have systematic errors caused by differences in the input physics of the grids; however, these are smaller than those caused by
uncertainties in $T_{\rm eff}$, [M/H], $\Delta\nu$ and $\nu_{\rm max}$ (Basu et al. 2012). 
Mathur et al. (2012) have shown that grid-based estimates of stellar masses and radii 
agree very well with those obtained from more detailed modeling of the oscillation frequencies, giving us
confidence in the robustness of our mass and radius estimates.
While detailed modeling of the oscillation spectrum is preferable for all stars, this is beyond the scope of this
paper.

\section{Results and Discussion}
\label{sec:results}

\begin{figure}
\epsscale{1.0}
\plotone{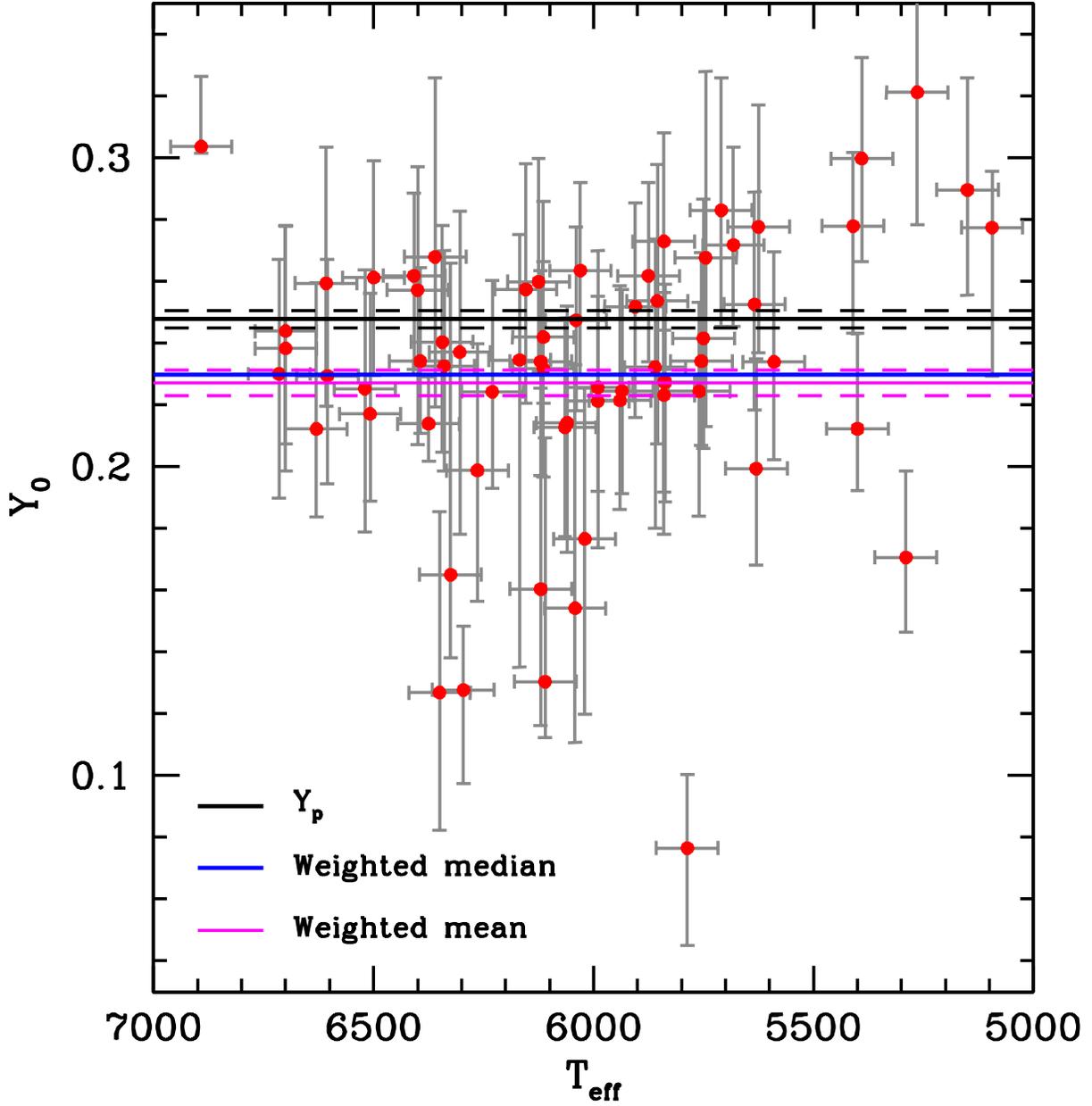}
\caption{The initial helium abundance, $Y_0$, of  stars in our sample obtained with the solar value of $\alpha$. { The
black line marks the primordial Big Bang nucleosynthesis helium fraction, $Y_p=0.2477\pm 0.0029$.
The pink line is the weighted mean of the sample, and the blue is the weighted median. The dashed lines show $1\sigma$ errors.}}
\label{fig:yt}
\end{figure}

The initial helium abundance, $Y_0$, needed to construct models of the stars in our sample --- assuming that they
all have the solar value of $\alpha$ --- is shown in Fig.~\ref{fig:yt}. Note that for $> 50$\% of our sample the
$Y_0$ estimate is  less than the primordial value of $Y_{\rm p}=0.2477\pm 0.0029$ (Peimbert et al. 2007).
{ While the deficit is within $1\sigma$ for some stars, there are stars with a $>3\sigma$ deficit. The
 weighted average of $Y_0$ of the sample is $0.227\pm 0.004$. The weighted median is $0.223$, both below $Y_{\rm p}$.}
It is, of course, highly unlikely that stars are born with less helium than was 
produced in the Big Bang.
Given that the only parameter we could change in MLT is $\alpha$, this implies
that solar $\alpha$ does not properly approximate convective heat transport in these stars.
Note that a change of physics inputs to the models will change the solar value of $\alpha$, and
the exercise repeated with the new solar $\alpha$ would give similar results.

\begin{figure}
\epsscale{0.8}
\plotone{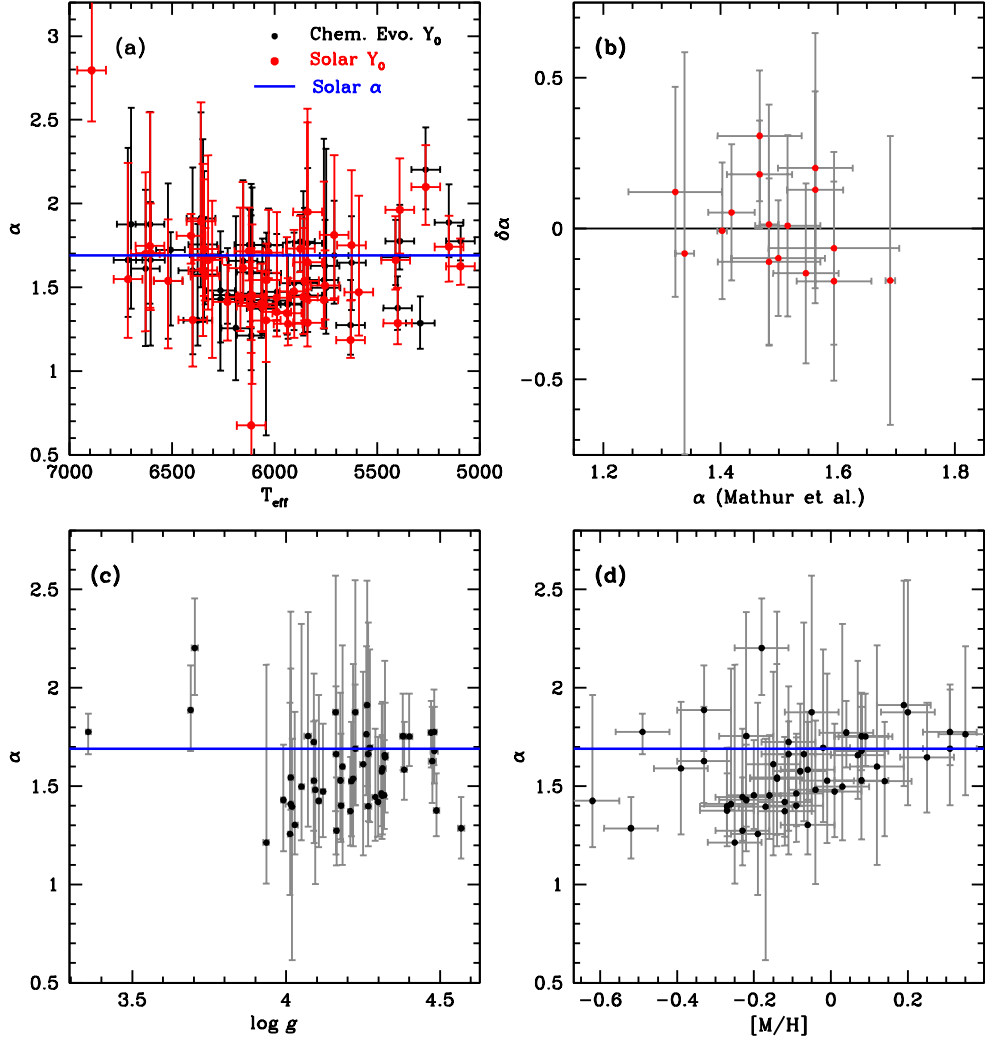}
\caption{(a) The derived mixing length  assuming the solar value of $Y_0$ (red) and for the 
chemical evolution model of $T_0$ (black). The blue line marks  solar $\alpha$. 
{
(b) The differences between $\alpha$ obtained by Mathur et al.~(2012) by fitting individual frequencies
and our $\alpha$ estimates of the same stars.
(c) $\alpha$ plotted as a function of $\log g$, and (d) $\alpha$ plotted as a function of [M/H]. Only $\alpha$
estimated with the chemical evolution model of $Y_0$ are plotted in (c) and (d).
}
}
\label{fig:at}
\end{figure}

In Fig.~\ref{fig:at}(a) we show the value of $\alpha$ for our sample obtained assuming
either the solar value of $Y_0$ (red points), or the simple chemical evolution model of $Y_0$ 
(black points).
Note that $\alpha$ for most of the stars is less than the solar value for both cases
 The average value of $\alpha$ for this sample is $1.522$ for solar $Y_0$ and
$1.597$ when the chemical-evolution model is used. 
In MLT,  lower $\alpha$ implies less efficient convection. Thus MLT
predicts that in the superadiabatic layers, convective energy transport in 
our sample is generally less efficient than that in the Sun.
Since the results with the two choices of $Y_0$ are similar, in the subsequent discussions
we only use $\alpha$ obtained with the chemical evolution model of $Y_0$.
{ 
Mathur et al. (2012) constructed detailed models to fit the mode frequencies of 
22 {\it Kepler} stars using the Asteroseismic Modeling Portal (AMP; Metcalfe et al. 2009) 
There are 16 stars in
common with our sample. In Fig.~\ref{fig:at}(b) we show the differences
 between the Mathur et al. $\alpha$ values and the ones obtained in this work.
The two $\alpha$ estimates agree well, mostly within $1\sigma$.
 Since the physics in the AMP
models is different from ours, they obtain a  solar $\alpha$ of 2.12. Thus to compare their results with
ours, we have scaled the AMP results to our value of the  solar $\alpha$. 
In Fig~\ref{fig:at}(c) and (d) we show the variation of  $\alpha$ with $\log g$ and [M/H].
}

{
In order to explore whether our $\alpha$ estimates are correlated with stellar properties,
we first determined the simple Spearman rank correlation between $\alpha$ and different
properties. The correlation coefficients are listed in Table~1.
Also listed is the  $p$-value, which is the  probability that the
correlation is a chance occurrence. A small $p$ therefore indicates a significant correlation.
There thus appears to be significant correlation between metallicity and
$\alpha$. There also seems to be a mildly significant correlation between mass and $\alpha$.
The significance of the correlation of $\alpha$ with $\log g$ or $T_{\rm eff}$
depends on whether or not we include the low-$\log g$ stars in our analysis. Table~1 lists
the coefficient obtained for the entire sample, as well as that obtained by removing 
 the lowest $\log g$ stars ($\log g < 3.8$) in our sample.
}

\begin{figure*}
\epsscale{0.68}
\plotone{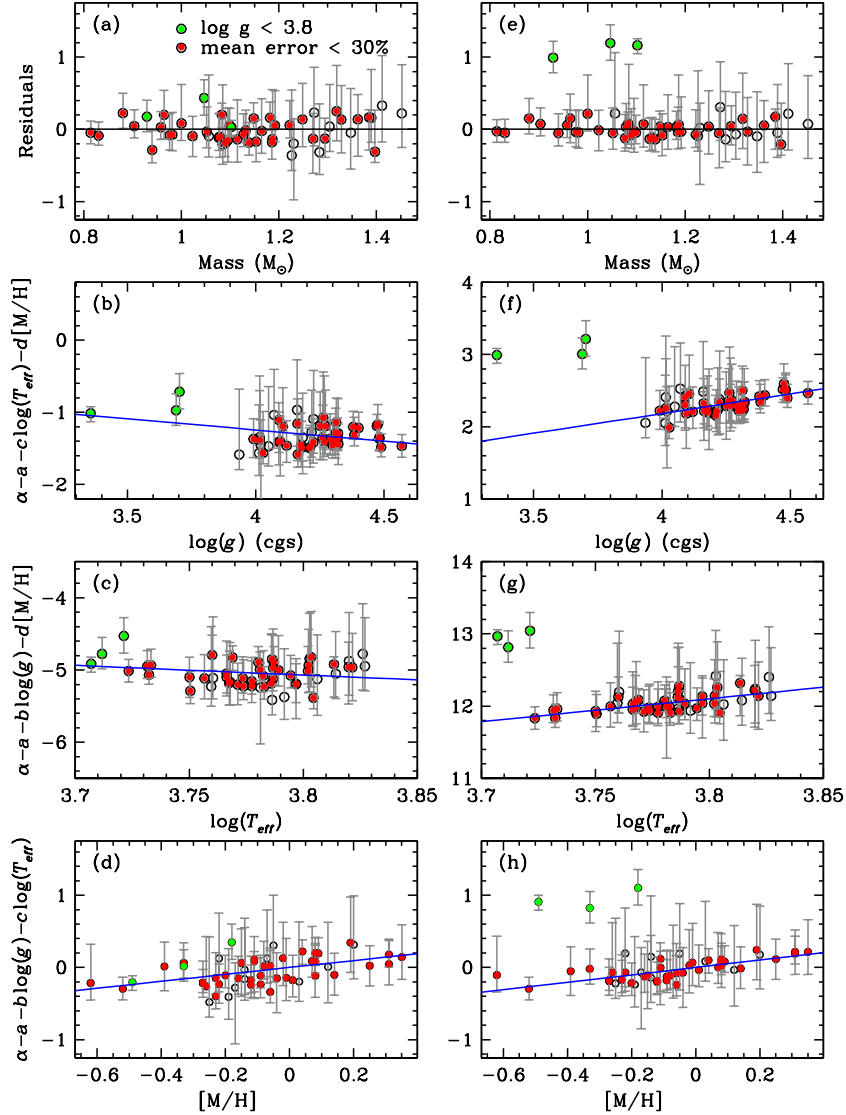}
\caption{
{ 
The residuals and partial residuals for the trilinear fit (Eq.~\ref{eq:mod}). Panels (a)-(d) are results when the entire
sample is used. Panels (e)-(h) are results for $\log g \ge 3.8$; the low $\log g$ points are however
still shown. The intercept $a$ changes when we apply the $\log g$ cutoff.
The blue lines are $b\log g$ [panels (b), (f)], $c\log T_{\rm eff}$ [Panels (c),(g)] and
$d$[M/H] [panels (d),(h)]. Note that the $\log g$ cut-off makes the correlations
tighter and that the [M/H] relation remains almost unchanged with and without the
$\log g$ cutoff.
}
}
\label{fig:avar}
\end{figure*}

{
Since $\alpha$ depends simultaneously on a number of parameters, to get a better estimate of
the correlations we perform a trilinear fit to $\alpha$ with the model
\begin{equation}
\alpha=a+b\log g+c\log T_{\rm eff}+d{\rm [M/H]}.
\label{eq:mod}
\end{equation}
Table~1 lists the coefficients and $p$-values, and  Fig.~\ref{fig:avar} shows the residuals, and partial
residuals, of the fit to Eq.~\ref{eq:mod}.
Note that the metallicity dependence is robust. The $\log g$ and $T_{\rm eff}$ correlations are
small and less statistically significant when the entire sample is used; these increase in
significance, but  change signs, when the $\log g$ cut is applied. The mass correlation seen 
in the Spearman correlation is most likely to be the result of the mass-$T_{\rm eff}$ and
mass-$\log g$ correlation seen in Fig.~1 as indicated by the fact that the residuals of the
fit to Eq.~\ref{eq:mod} do not show any trend with mass (Fig.~\ref{fig:avar}(a) and (e)).
}

{ The metallicity dependence of $\alpha$ is relatively easy to understand. It is most likely
caused by the temperature sensitivity of the H$^-$ density and hence, of its dominant
contribution to the optical continuum opacity. 
E.g., in the solar photosphere, $\sim 50$\% of the electrons that form H$^-$ are donated by metals,
and the fraction increases steeply with height due to low-ionization potential
elements like Na, Al, K, Ca and Cr.  With a smaller amount of metals, the temperature 
sensitivity of the H$^-$ density will therefore increase. 
This in turn will increase the contrast between up- and down-flows, and especially
increase the range of depths over which the down-flows will be cooled.
This results in a larger entropy jump between the surface and the deeper
layers, meaning a lower convective efficiency (a smaller $\alpha$ in the
context of MLT).
}
Since the average metallicity of our sample is sub-solar, we believe that this  metallicity dependence accounts 
for the lower-than-solar average value of $\alpha$ for our sample.

The lack of a significant correlation between $\alpha$ and  $T_{\rm eff}$ or $\log g$ 
is surprising.  This is most likely the  result of the limited and skewed range of $\log g$ of 
our sample,  and is confirmed by the change of the sign of the correlation when the $\log g$ cutoff is applied.
 A larger sample should resolve these issues, in particular, data on giants should help determine the
$\log g$ dependence properly. 
Piau et al. (2011), using a sample of red giants with
radii known from interferometry, have shown that the  red-giant models require sub-solar values of $\alpha$ to 
explain the observations; however, they did not address the  dependence of $\alpha$ on stellar parameters.

As noted earlier, $\alpha$ is just a proxy  for describing convection in stars. 
Although it is known that using such a proxy does not  reproduce properties of the stellar near-surface layers
correctly, MLT remains a practical tool in stellar modeling. 
An $\alpha$ type parameter can also be derived from numerical simulations 
of stellar convection (e.g., Ludwig et al. 1999; Trampedach et al. 1999; Trampedach 2007).
At present it is difficult to compare these  results with ours since the
simulations were  for solar composition, and the metallicity dependence of
our results is fairly strong. However, there do seem to be some
differences between our findings  and the simulations. At a given
$\log g$,  {  Ludwig et al. (1999)  found  $\alpha$ to decrease with increasing  $T_{\rm eff}$ for dwarfs in their 2D 
simulations and they find $\alpha$ to decrease with $\log g$.  A similar behavior was seen in the 3D simulations of Trampedach (2007).
Our $\alpha$-$T_{\rm eff}$ correlation agrees with theirs, 
but the $\log g$ one does not when we examine the entire sample; when we apply the
 $\log g$ cutoff
the reverse becomes true, the $\log g$ correlation agrees, the $T_{\rm eff}$ correlation does not.
}

As more detailed asteroseismic data become available from {\it Kepler}, and they are modeled, we will
be able to reduce the uncertainties in the mixing-length parameters needed to model the stars, and the
dependence of the properties of near-surface convection, including the corresponding value of $\alpha$, 
on stellar parameters will become clearer. As it is, our results have important 
implications for the different branches of astrophysics that depend on fitting stellar models. 
The usual way to determine stellar properties is through spectroscopic
or photometric analyses of the star combined with fitting to grids of
stellar models to obtain masses and radii (e.g., Takeda et al. 2007) or ages (e.g., J{\o}rgensen \& Lindegren 2005).
At a given metallicity, a lower $\alpha$ makes the evolutionary track of a given mass redder 
than its higher-$\alpha$ counterpart, and thus the properties of a star obtained using models constructed with a solar $\alpha$
will be quite different from those obtained with a sub-solar $\alpha$. Although the $T_{\rm eff}$ change
with $\alpha$ is really a result of a radius change, it appears as a change in the estimated mass
of star being fitted (Basu et al. 2011).  Consequently, we would be underestimating the mass 
of a sub-solar metallicity star if we use models constructed with the solar value of $\alpha$. Stellar population and
spectral synthesis models also use a single (usually the solar calibrated) value of $\alpha$ (see e.g. Coelho et al. 2007),
and our results now show that the uncertainties in $\alpha$ need to be added to the error budget of results that
use those models.

\acknowledgments
We thank the anonymous referee for comments that have improved this paper.
Funding for the {\it Kepler} mission is provided by NASA's Science Mission Directorate.
This work was partially supported by NSF grant AST-1105930 and NASA grant NNX09AJ53G.
We also thank all other funding councils and agencies
that have supported the activities of  Working Group 1 of the {\it Kepler} Asteroseismic Science 
Consortium.

\newpage

\begin{table}
\caption{Correlations and the associated $p$ value between $\alpha$ and stellar parameters.}
\begin{center}
{\small
\begin{tabular}{lcclcc}
\hline
\noalign{\smallskip}
 &\multicolumn{5}{c}{Spearman Correlation}\\
{\ } &\multicolumn{2}{c}{All data} &{\ } & \multicolumn{2}{c}{$\log g \ge 3.8$} \\
Parameter &  $r$ & $p$-value &
          & $r$ & $p$-value \\
\noalign{\smallskip}
\hline
\noalign{\smallskip}
$\log g$             &  $0.110$  & 0.436 & & 0.316 & 0.027  \\
$\log(T_{\rm eff})$  &  $0.039$  & 0.781 & & 0.224 & 0.122  \\
{[M/H]}              &  $0.445$  & 0.001 & & 0.636 & $10^{-5}$ \\
Mass                 &  $0.154$  & 0.274 & & 0.263 & 0.067  \\
\noalign{\smallskip}
\hline
\noalign{\smallskip}
 &\multicolumn{5}{c}{Trilinear Analysis}\\
{\ } &\multicolumn{2}{c}{All data} &{\ } & \multicolumn{2}{c}{$\log g \ge 3.8$} \\
& fitted-value & $p$ & & fitted-value & $p$ \\
\noalign{\smallskip}
\hline
$a$ & $7.97\pm 0.27$ & 0.010 & & $-12.77\pm2.91$ & $6.8\times10^{-5}$\\
$b$ & $-0.31\pm 0.09$ & 0.002 & & $0.54\pm0.11$ & $1.7\times10^{-5}$\\
$c$ & $-1.33\pm 0.80$ & 0.102 & & $3.18\pm0.69$ & $3.3\times10^{-5}$\\
$d$ & $0.48\pm 0.12$ & $2\times10^{-3}$& & $0.52\pm 0.07$ & $4.5\times10^{-9}$\\
\noalign{\smallskip}
\hline
\end{tabular}
}
\end{center}
\label{tab:corr}
\end{table}

\newpage

\end{document}